\documentclass[prd,aps,showpacs,nofootinbib,floats]{revtex4}
\usepackage{amsmath,amssymb}
\usepackage{graphicx}

\usepackage{bm}

\newcommand{\myodot}{{\mathchoice
    {\raisebox{\depth}{$\displaystyle\odot$}}
    {\raisebox{\depth}{$\textstyle\odot$}}
    {\raisebox{\depth}{$\scriptscriptstyle\odot$}}
    {\raisebox{\depth}{$\scriptscriptstyle\odot$}}
    }}

\begin{document}

\bibliographystyle{apsrev}

\title[Short Title]{Revising Limits on Neutrino-Majoron Couplings}

\author{A. P. Lessa}\email{andlessa@ifi.unicamp.br}
\author{O. L. G. Peres}\email{orlando@ifi.unicamp.br}

\affiliation{
  Instituto de F\'\i sica Gleb Wataghin - UNICAMP,
  13083-970, C.P. 6165, Campinas SP, Brazil
}

\newcommand{\lsim}{\,\lower .5ex\hbox{$\buildrel < \over {\sim}$}\,}
\newcommand{\gsim}{\,\lower .5ex\hbox{$\buildrel > \over {\sim}$}\,}
\newcommand{\vect}[1]{\overrightarrow{\sf #1}}
\newcommand{\system}[1]{\left\{\matrix{#1}\right.}
\newcommand{\displayfrac}[2]{\frac{\displaystyle #1}{\displaystyle #2}}
\newcommand{\nucl}[2]{{}^{#1}\mbox{#2}}
\newcommand{\diff}{{\rm\,d}}
\newcommand{\ea}{{\em et al.}}
\newcommand{\be}{\begin{equation}}
\newcommand{\ee}{\end{equation}}

\begin{abstract}
Any theory that have a global spontaneously broken symmetry will
imply the existence of very light neutral bosons or massless bosons
(sometimes called Majorons). For most of these models we have
neutrino-Majoron couplings, that appear as additional branching
ratios in decays of mesons and leptons. Here we present an updated
limits on the couplings between the electron, muon and tau neutrinos
and Majorons. For such we analyze the possible effects of Majoron
emission in both meson and lepton decays. In the latter we also
include an analysis of the muon decay spectrum. Our results are
 $|g_{e\alpha}|^{2}<5.5\times10^{-6}$,
$|g_{\mu\alpha}|^{2}<4.5\times10^{-5}$ and
$|g_{\tau\alpha}|^{2}<5.5\times10^{-2}$ at 90 \% C. L., where
$\alpha=e,\mu,\tau$.
\end{abstract}

\maketitle

\baselineskip=20pt

\section{Introduction}

\label{intro}

Recently neutrino physics has given us many surprises with strong
evidences for flavor neutrino conversion to another type of
neutrinos. Analysis of data from solar, atmospheric and reactor
neutrinos have shown us that no other mechanism can explain all the
data, unless you have massive
neutrinos~\cite{Mohapatra:2006gs,Smirnov:2007pw}. These experiments
are the first strong evidence for non-conservation of
family lepton number and this may indicate that new symmetries and
interactions are the source of this phenomena.

Experimental evidences of massive neutrinos imply that the Minimal
Standard Model (SM) is no longer correct. The simplest extension would
be the inclusion of right-handed sterile neutrinos, what would allow
Dirac mass terms for neutrinos. Despite its simplicity, this approach
does not help us to understand the neutrino mass scale or predict
the neutrino masses. Due to the large gap between neutrino mass scales
and the other SM scales, several mechanisms have been suggested to
generate neutrino masses, relating this mass scale to new physics.
In many of these models the masses are of the Majorana type or a mix
between Majorana and Dirac types, what implies non-conservation of
lepton number.

As it's well known, lepton number is an accidental global symmetry
($U_{L}(1)$) of the Standard Model. So, if the neutrino mass matrix
includes Majorana terms, lepton number is broken either explicitly
or spontaneously. If lepton number ($L$) is indeed a global
symmetry\footnote{In some grand unified theories (GUTs), lepton number
is gauged and becomes a subgroup of a larger gauge symmetry.%
}, its spontaneous breaking will generate a Goldstone boson, usually
called Majoron~\cite{Mohapatra, Gelmini}. In this case the breaking of
$L$ sets a new scale and
requires a scalar which carries lepton number and acquires a non-null
vacuum expectation value (vev). Several extensions of the SM allow
spontaneous $L$ breaking
and predicts the existence of the Majoron. However, the simplest extensions
(with a triplet scalar) are excluded due to the experimental results
of LEP on the $Z^{0}$ invisible decay.

Another important class of models which predicts the existence of
Majorons are supersymmetric extensions of the Standard Model with
spontaneous R parity breaking. In these models, the introduction of
anti-neutrino superfields ($N^{C}$) and new singlet superfields
($\Phi$) (which contain neutral leptonic scalars), allows
spontaneous breaking of lepton number~\cite{Masiero, Romao, Romao2}.
In almost all of these models the Majoron will be the imaginary part
of some linear combination of sneutrinos, the scalar component of
the super Higgs fields ($H_{u}$ and $H_{d}$) and the $\Phi$
superfields. Therefore we may safely assume as a model-independent
coupling the following interaction term between $J$ and $\nu$ %
\footnote{The same being valid for non-supersymmetric models as well. %
}:
\begin{eqnarray}
\mathcal{L}=\sum_{\alpha,\beta=e,\mu,\tau}
ig_{\alpha\beta}\bar{\nu}_{\alpha}\gamma^{5}\nu_{\beta}J\mbox{, }
\end{eqnarray}
where $g_{\alpha\beta}$ is a general complex coupling matrix in the
flavor basis. Because in most models $J$ is basically a singlet
(avoiding the constraints imposed by the LEP results), the above
couplings are usually the most
relevant ones to phenomenological analysis (at least at low energies).
In most models we must also include couplings between neutrinos
and a new light scalar (that we call $\chi$) with the same couplings as $J$:
\begin{eqnarray}
\mathcal{L}=\sum_{\alpha,\beta=e,\mu,\tau}
g_{\alpha\beta}\bar{\nu}_{\alpha}\nu_{\beta}\chi\mbox{,}
\end{eqnarray}
Usually neutrino masses and mixings will depend on the vevs associated
to the spontaneous breaking of $L$ and the matrix $g$. In this
context, knowledge of the couplings between neutrinos and Majorons
may help us to understand the neutrino mass scale. However, this relation
is very model dependent and may be very hard to realize in practice.
Trying to make our results as model independent as possible, we will
make no assumptions on $g_{\alpha\beta}$ and present our results
with and without the existence of the massive scalar $\chi$. Nevertheless, assuming Majorana neutrinos (what is reasonable since lepton number is violated), bounds on $g_{\alpha \beta}$ may be transformed to the mass basis through the relation
\begin{equation}
G=U^{T}gU
\end{equation}
where $G_{ij}$ is the neutrino-Majoron coupling matrix in the basis where the neutrino mass matrix is diagonal ($M=diag(m_{1},m_{2},m_{3})$) and $U$ rotates the mass eigenstates to the flavor eigentates (see section \ref{iid}).

Majoron models can be interesting from the cosmological point of
view because they can  affect bounds to neutrino masses from large
scale structure~\cite{beacom1}. Neutrinos coming from astrophysical
sources can also be significantly affected by fast decays, where
the only mechanism not yet eliminated is due to neutrinos coupling
to Majorons. This can affect the very high energy region, strongly
changing the flavor ratios between different neutrino
species~\cite{beacom2}  or the lower energy region, as supernova
neutrinos~\cite{yasaman1,tomas2}.
\begin{table}
\begin{tabular}{|c|c|c|c|}
\hline Category& Upper Bound& Process & Reference \tabularnewline
\hline \hline
\hline
& & &  \tabularnewline
solar neutrino constrain& $|G_{21}|^2 <  4\times 10^{-6}
\left(\dfrac{7\times  10^{-5} ~{\mbox eV}^2}{\Delta
m^2_{\myodot}}\right)$& $\nu_{2} \rightarrow J + \nu_{1}$&
\cite{beacom3}\tabularnewline
& & &  \tabularnewline
& & &  \tabularnewline  \hline
& & &  \tabularnewline
supernova bounds & $|g_{ee}| < (1-20) \times 10^{-5}
\mbox{,} \quad 2  \times 10^{-11}<
|g_{e\mu}||g_{\mu\mu}| < 3   \times 10^{-10}$& $\nu
\rightarrow J+\nu\mbox{, }\nu+\nu\rightarrow J$& \cite{yasaman1}
\tabularnewline
& $|g_{\alpha\beta}| < 3  \times 10^{-7}\mbox{ or } |g_{\alpha\beta}|
> 2 \times 10^{-5}$& $\nu
\rightarrow J+\nu^{\prime}\mbox{, }\nu+\nu\rightarrow J$& \cite{tomas2} \tabularnewline
& & &  \tabularnewline
& & &  \tabularnewline  \hline
& & &  \tabularnewline
$\beta\beta 0\nu$ decay & $|g_{ee}| < 2 \times
10^{-4}$ &
$(A,Z)\rightarrow (A,Z+2)+2e+J$&
\cite{dbeta} \tabularnewline
& & &  \tabularnewline
& & &  \tabularnewline  \hline
& & &  \tabularnewline
microwave background data & $G_{ij} \leq \,  0.61\, 10^{-11}\, m_{50}^{-2}\quad  {\mbox and
}\quad   G_{ii} \leq 10^{-7}$  & $\nu
\rightarrow J+\nu^{\prime}$ & \cite{Hannestad:2005ex} 
 \tabularnewline
& & &  \tabularnewline
& & &  \tabularnewline  \hline
& & &  \tabularnewline
meson decay &
$\sum_{l=e,\mu,\tau}|g_{el}|^2 < 3 \times 10^{-5} \mbox{,} \quad
\sum_{l=e,\mu,\tau} |g_{\mu l}|^2 < 2 \times 10^{-4}$
&
$\pi /K\rightarrow e+\nu +J$& ~\cite{Barger,Gelmini2,Britton:1993cj} \tabularnewline
& & &  \tabularnewline
& & &  \tabularnewline  \hline
\end{tabular}
\caption{Some of the previous bounds on neutrino-Majoron couplings
from different sources. In the last two columns are shown the process
used to constraint the couplings and the respective references.}
\label{Tab0}
\end{table}

Presently we know that the role of neutrino-Majoron couplings is
marginal in solar and atmospheric neutrinos, therefore it's possible
to put a limit on~\cite{beacom3}
\begin{eqnarray}
|G_{21}|^2 < | g_{\alpha\beta} U^{*}_{\alpha 2}U_{\beta 1}|^2 <
4\times 10^{-6} \left(\dfrac{7\times  10^{-5} ~{\mbox eV}^2}{\Delta
m^2_{\myodot}}\right)
\end{eqnarray}
where $G_{21}$ is the neutrino-Majoron coupling in the mass basis
and $\Delta m^2_{\myodot}$ is the solar mass difference squared
($\Delta m^2_{21}~\equiv~ m^2_{2}-m^2_{1}$). The observation of
1987A explosion ensures us that a large part of binding energy of
supernova is released into neutrinos, what can be translated into
the bounds~\cite{yasaman1}
\begin{eqnarray}
|g_{ee}| < (1-20) \times 10^{-5}\mbox{,} \quad 2  \times 10^{-11}<
|g_{e\mu}||g_{\mu\mu}| < 3   \times 10^{-10}
\end{eqnarray}
such bounds where read off from Fig.1 and Figs. 3,4 of
Ref.~\cite{yasaman1} for  $g_{ee}$ and $|g_{e\mu}||g_{\mu\mu}|$,
respectively. Also, limits from decay and scattering of Majorons
inside  supernova give the bounds~\cite{tomas2}
\begin{eqnarray}
|g_{\alpha\beta}| < 3  \times 10^{-7}\mbox{ or } |g_{\alpha\beta}|
> 2 \times 10^{-5}.
\end{eqnarray}
the first limit appears because if neutrino-Majoron coupling is
strong enough the supernova energy is drained due Majoron emission
and no explosion occurs; the second limit appears  because if
neutrino-Majoron coupling is too strong, the Majoron becomes trapped
inside the supernova and no constraint is possible.

While neutrinoless double beta decays ($\beta\beta 0\nu$) provide us the constraint
\begin{eqnarray}
|g_{ee}| < 2 \times 10^{-4}\mbox{ and } |g_{ee}| < 1.5
\end{eqnarray}
where the first (second) bound corresponds to Majorons with lepton
number equal to L=0 (L=2) at 90 \% C. L.~\cite{dbeta}. Also, no evidence of
Majoron production was seen in pion and kaon decays and
therefore~\cite{Barger,Gelmini2,Britton:1993cj}
\begin{eqnarray}
\sum_{l=e,\mu,\tau} |g_{el}|^2 < 3 \times 10^{-5} \mbox{,} \quad
\sum_{l=e,\mu,\tau} |g_{\mu l}|^2 < 2 \times 10^{-4}.
\end{eqnarray}
Besides the bounds mentioned above, there are bounds that depend on
the rate of neutrino decay ($\nu \rightarrow \nu^{\prime} J$). Such
reaction depends on the neutrino lifetime, $\tau$, that is a
function of neutrino-Majoron couplings in the mass basis, which we
denote by G. Without additional assumptions on neutrino hierarchy,
we can not relate directly the neutrino-Majoron couplings and the
neutrino lifetime.  One  example is Ref.~\cite{Hannestad:2005ex}
that using cosmic microwave background data, put a stronger
constrain
\begin{equation}
G_{ij} \leq \,  0.61\, 10^{-11}\, m_{50}^{-2}\quad  {\mbox and
}\quad   G_{ii} \leq 10^{-7} \label{hann-raff}
\end{equation}
where $m_{50}=m/50\;meV$ and G is the neutrino-Majoron in the mass
basis: $G_{ii}$ and $G_{ij}$ are respectively the diagonal and
off-diagonal elements of G.

Future experiments can improve the present bounds on many order of
magnitude, we refer to Ref.~\cite{Fogli:2004gy,Serpico:2007pt} for
details.

A summary of some of the previous bounds are shown in
Table~\ref{Tab0}, where we also show the respective relevant process
used to constraint the neutrino-Majoron couplings. Almost all the
bounds shown in Table~\ref{Tab0} assume one particular model or
class of models. Probably the most model-independent result is
from~\cite{Barger,Gelmini2,Britton:1993cj}, but in this case they
assume not only neutrino-Majoron couplings but also neutrino-$\chi$
couplings to compute the upper bounds shown in Table~\ref{Tab0}.

Here we will try to improve or make these limits more
model-independent through an analysis of  both meson and
lepton decays. In Section \ref{iia} we discuss the limits from
pion, kaons, D, D$_s$ and B decays, including decays of mesons
into taus; in Sections \ref{iib} and \ref{iic} we include bounds from
lepton decays (from the total rate and from the spectral distortions).
We conclude transforming our bounds to the mass basis in Section \ref{iid}.

\section{Results}
\label{ii}

Here we try to improve the bounds on neutrino-Majoron couplings through
the analysis of possible effects on mesons and leptons decays as well
as on the spectrum of the muon decay. We also rewrite our results
in the mass basis, which in many cases is more important for model
analysis. All the bounds obtained here have 90\% C.L. and were obtained through
the chi-square method assuming gaussian distributions and including
both statistical and theoretical errors as follows
\begin{equation}
\chi^{2}=\dfrac{\left(R_{data}-R_{theor} \right)^2}{\sigma^2_{data}+\sigma^2_{theor}}
\label{chi}
\end{equation}
where $R_{data}$, $R_{theor}$, $\sigma_{data}$ and $\sigma_{theor}$
are respectively the experimental data of the rate R, the theoretical prediction
for process R, assuming an  incoherent sum of SM rate and Majoron contribution, the
experimental error and the theoretical error.

\subsection{Meson decay rates}
\label{iia}
The process $M\rightarrow l+\nu_{l}$ was extensively studied in the
literature and has the following total
decay rate~\cite{PDG}:
\begin{equation}
\Gamma_{SM}=\dfrac{G_{F}^{2}|V_{qq'}|^{2}}{8\pi}f_{m}^{2}m_{l}^{2}m_{M}
\left(1-\dfrac{m_{l}^{2}}{m_{M}^{2}}\right)^{2} f_{rad},
\label{MesonDecay}
\end{equation}
where the $f_{rad}$ accounts for radiative corrections. In
Eq.(\ref{MesonDecay}), $m_{M}$ and $m_{l}$ are the meson and lepton masses,
$G_{F}$ is the Fermi constant, $f_{m}$ is the
meson decay constant and $V_{qq'}$ is the respective
Cabibbo-Kobayashi-Maskawa (CKM) matrix element. Unless specified otherwise
we are using the quantities as listed in the Particle Data group
compilation~\cite{PDG}. We also use the same source to compute the relevant  radiative
corrections for the  mesons decay rates. An important
feature of Eq.(\ref{MesonDecay}) is that, because it's a 2-body decay,
$\Gamma_{SM}$ is proportional to $m_{l}^{2}$, as it should be to conserve
angular momentum.

In the last few years several of the meson decay constants were
calculated through lattice QCD~\cite{LatticeQCD}, which can be used
to obtain stronger theoretical predictions. We used both the
experimental~\cite{PDG} and theoretical
values~\cite{Blattice,Kdecaycte,Dsdecaycte1,Dsdecaycte2} of $f_{m}$
on our calculations, but in most cases the results differ only by 10\%.
For this reason we will only show the results using the experimental
values of $f_{m}$.

Due to the neutrino-Majoron couplings, the following process also
contributes to mesons decay rates:
\begin{eqnarray}
M\rightarrow l+\nu_{l'}+J,
\end{eqnarray}
where $J$ stands for Majoron and $\nu_{l'}$ may be any neutrino
flavor. A complex analytic expression for the total decay rate is
given in~\cite{Gelmini2}. Here we show a simpler result valid in the limit
$m_{l}=m_{\nu}=0$:
\begin{eqnarray}
\Gamma_{J}=\dfrac{G_{F}^{2}|V_{qq'}|^{2}}{768\pi^{3}}f_{m}^{2}m_{M}^{3}
\sum_{m=e,\mu\tau}|g_{l m}|^{2}
\end{eqnarray}
This result shows that when Majorons are included, the total decay
rate is no longer proportional to the lepton mass (since now we have
a 3-body decay). Therefore, the Majoron contribution ($\Gamma_{J}$) may easily
overcome the SM predictions ($\Gamma_{SM}$) if $g\sim1$:
\begin{eqnarray}
\dfrac{\Gamma_{J}}{\Gamma_{SM}}\approx\dfrac{1}{48\pi^{2}}
\dfrac{m_{M}^{2}}{m_{l}^{2}}\gg 1
\end{eqnarray}
where we have assumed $m_{l} \ll m_{M}$. Assuming that the total decay rate is
\begin{eqnarray}
\Gamma_{total}=\Gamma_{SM}+\Gamma_{J}\label{TotalTax} ,
\end{eqnarray}
the decay on $J$ will be  the dominant channel, unless $g$ is
small.
Because only small deviations from the SM are allowed by experimental
data, we must have $g\ll1$.  Following Eq.(\ref{chi}),  we
calculated upper bounds for $|g_{\alpha \beta}|^2$ at 90 \% C. L. The
Table~\ref{Tab1} shows the bounds obtained through this analysis.
\begin{table}
\begin{tabular}{|c|c|c|c|c|}
\hline
Mesons&
$|g_{e\alpha}|^{2}$&
$|g_{\mu\alpha}|^{2}$&
$|g_{\tau\alpha}|^{2}$&
Refs (exp. values)
\tabularnewline
\hline
\hline
$\pi$&
$1.6\times10^{-4}$&
$2.1\times10^{-1}$&
&
\cite{PDG}
\tabularnewline
\hline
$K$&
$9.5\times10^{-4}$&
$9.3$&
&
\cite{PDG}
\tabularnewline
\hline
$D$&
$1.6\times10^{-1}$&
$2.3$&
$23$&
\cite{PDG}
\tabularnewline
\hline
$D_{s}$&
&
$1$&
$6.3$
&
\cite{Artuso:2006kz}
\tabularnewline
\hline
$B$&
$0.85$&
$1.5$&
$19$&
\cite{Satoyama:2006xn}
\tabularnewline
\hline
\end{tabular}
\caption{Upper bounds on $\sum_{l=e,\mu,\tau}|g_{l \beta}|^{2}$
 from meson decays with 90\% C.L. The references for the experimental values used are
shown in the last column. We only include the Majoron contribution,
and not the new light scalar $\chi$. }
\label{Tab1}
\end{table}
As expected from the above remarks and the results on Table
\ref{Tab1}, the most constrained matrix elements $g_{\alpha\beta}$
will be those concerning $e$, since the approximation $m_{l} \ll
m_{M}$ is good in this case. We found that this bound can be
improved using recent data~\cite{NA48} of the following ratio:
\begin{eqnarray}
\dfrac{\Gamma(K^{+}\rightarrow
e^{+}+\nu_{e})}{\Gamma(K^{+}\rightarrow\mu^{+}+\nu_{\mu})}=
(2.416\pm0.043)\times10^{-5}
\end{eqnarray}
where the error is the quadrature of statistical and systematic errors. Because
the Majoron contributions must be suppressed (as shown in Table \ref{Tab1}), we
may approximate the above ratio:
\begin{eqnarray}
\dfrac{\Gamma(K^{+}\rightarrow
e^{+}+\nu_{e})}{\Gamma(K^{+}\rightarrow\mu^{+}+\nu_{\mu})}=
\dfrac{\Gamma_{SM}^{e} + \Gamma_{J}^{e}}{\Gamma_{SM}^{\mu} + \Gamma_{J}^{\mu}}\approx
\dfrac{\Gamma_{SM}^{e} + \Gamma_{J}^{e}}{\Gamma_{SM}^{\mu}} ,
\end{eqnarray}
where $\Gamma^{e(\mu)}$ represents the decay rate with an $e$ ($\mu$)
in the final state. In this way, using the previous statistical
analysis, we can constraint the elements $g_{e\alpha}$ (at 90\% C.L.):
\begin{eqnarray}
\sum_{l=e,\mu,\tau}|g_{el}|^{2}<1.1\times10^{-5}
\end{eqnarray}
When it comes to the $\mu$ matrix elements ($g_{\mu\alpha}$), the
constraints from Table \ref{Tab1} may also be improved if we
consider the decay channels of mesons in four leptons~\cite{PDG}:
\begin{eqnarray}
BR(K^{+}\rightarrow\mu^{+}+\nu_{\mu}+\nu+\bar{\nu})<6\times10^{-6}
\end{eqnarray}
Since the SM contribution to this decay is negligible, we may assume:
\begin{eqnarray}
BR(K^{+}\rightarrow\mu^{+}+\nu_{l'}+J)<6\times10^{-6}
\end{eqnarray}
resulting on (at 90\% C.L.):
\begin{eqnarray}
\sum_{l=e,\mu,\tau}|g_{\mu l}|^{2}<9\times10^{-5}
\end{eqnarray}
Finally, new experimental data for leptonic decay rates of heavy
mesons such as the $D^{+}$, $D_{s}^{^{+}}$ and $B^{^{+}}$
mesons~\cite{PDG,Corwin:2006wb,Satoyama:2006xn,Artuso:2006kz,Artuso:2005ym},
allow us to impose limits to the $\tau$ matrix elements
($g_{\tau\alpha}$), as shown in Table \ref{Tab1}. The best bound being
from the $D_{s}^{+}$ leptonic decay on $\tau^{+}+\nu_{\tau}$ (at 90\% C.L.)\cite{Artuso:2006kz}:
\begin{eqnarray}
\sum_{l=e,\mu,\tau}|g_{\tau l}|^{2}<6.3
\end{eqnarray}
Because of large experimental uncertainty, this bound is quite weak,
as can be seen above.

We stress that unlike~\cite{Barger,Gelmini2,Britton:1993cj} the results shown
so far do not include possible decays on a light scalar $\chi$ and
therefore are less model-dependent. If this new scalar is considered
with a mass of 1 KeV (other choices for the $\chi$ mass
do not change these results as long as it is well below the initial state
masses), the previous results are basically improved by a factor of 2 (again, at 90\% C.L.):
\begin{eqnarray}
\sum_{\alpha}|g_{e\alpha}|^{2}<5.5\times10^{-6}\mbox{ ,
}\sum_{\alpha}|g_{\mu\alpha}|^{2}<4.5\times10^{-5}\mbox{
and }\sum_{\alpha}|g_{\tau\alpha}|^{2}<3.2
\label{boundmeson}
\end{eqnarray}

\subsection{Lepton decay rates}
\label{iib}
Because of its good experimental precision, lepton
decays are good candidates for imposing bounds on neutrino-Majoron
couplings. Moreover, in this case there aren't uncertainties such as
mesons decay constants and CKM elements. However, the leading term
in $\Gamma(l_{i}\rightarrow l_{j}+\bar{\nu}_{j}+\nu_{i})$ is no
longer proportional to the final lepton mass (as it was in the case
of mesons), because the SM decay is a 3-body decay already. For this
reason $\Gamma_{J}<\Gamma_{SM}$ even for $g\sim 1$. In fact the
inclusion of Majorons in the final state decreases the decay rate by
a factor of $\approx 10$ ($\Gamma_{J} \approx \Gamma_{SM}/10$, for
$g=1$), instead of increasing it as it was in the case of mesons.
Therefore we expect much weaker bounds in this case. But, as we will
show below, it's still possible to obtain good bounds for certain
decays\footnote{We thanks J. F. Beacom for suggestion to use lepton
decays to constrain neutrino-Majoron decays.}. To calculate the
4-body decay rate~($\Gamma(l\rightarrow l'+\bar{\nu}+\nu+J$)) we
used the programs FeynArts and FormCalc~\cite{Hahn:2000kx,Hahn:1998yk}.

As we did in the meson case, to constraint the $g_{\alpha\beta}$
matrix we assume that the total lepton decay rate receives
contributions from Majoron emission:
\begin{eqnarray}
\Gamma_{total}(l_{\alpha}\rightarrow
l_{\beta}+\bar{\nu}_{\beta}+\nu_{\alpha})=\Gamma_{J}(l_{\alpha}\rightarrow
l_{\beta}+\bar{\nu}+\nu+J)+\Gamma_{SM}(l_{\alpha}\rightarrow
l_{\beta}+\bar{\nu}_{\beta}+\nu_{\alpha})
\end{eqnarray}
Because Majoron emission may change neutrino flavor (since
$g_{\alpha\beta}$ may be non-diagonal), $\Gamma_{J}$ may have any
type of neutrinos in its final state. For this reason we omitted the
subindex in $\Gamma_{J}$. Besides, both neutrinos ($\nu$ or
$\bar{\nu}$) may emit Majorons, what implies:
\begin{eqnarray}
\Gamma_{J}(l_{\alpha}\rightarrow l_{\beta}+\bar{\nu}+\nu+J)\propto
\sum_{\delta}(|g_{\alpha \delta}|^{2}+|g_{\beta \delta}|^{2})
\label{LepProp}
\end{eqnarray}
where $g_{\alpha \delta}$ and $g_{\beta \delta}$ are the couplings
between Majoron and the $\alpha$ anti-neutrino and $\beta$ neutrino,
respectively. In Eq.(\ref{LepProp}), the interference terms
$g_{\alpha \delta}g_{\beta \delta}$ are proportional to neutrino
masses squared and were neglected. Because Table~\ref{Tab1} shows
that lighter leptons have stronger upper bounds, we will assume
$g_{\alpha\delta}\gg g_{\beta\delta}$. Therefore we will consider
that Majoron emission by $\bar{\nu}_{\alpha}$ is dominant:
\begin{eqnarray}
\Gamma_{J}(l_{\alpha}\rightarrow l_{\beta}+\bar{\nu}+\nu+J)\propto
\sum_{\delta}|g_{\alpha \delta}|^{2}
\end{eqnarray}

Using the experimental values for the $\mu$ and $\tau$ decay
rates~\cite{PDG} and the same kind of analysis used in the last
section, the following bounds were obtained at 90\% C.L.:
\begin{eqnarray}
\sum_{\alpha}|g_{\mu\alpha}|^{2}<4\times 10^{-4}\mbox{ ,}\quad
\sum_{\alpha}|g_{\tau\alpha}|^{2}<10\times 10^{-2},
\end{eqnarray}
where the first bound comes from $\mu$ decay and the second from
$\tau$ decay, both at 90\% C.L.. For the $\tau$ decay the same constraint is
obtained if one considers decays in $e$'s or $\mu$'s. If we include
the contributions from $\chi$ emission (again with mass of 1 KeV and at 90\% C.L.):
\begin{eqnarray}
\sum_{\alpha}|g_{\mu\alpha}|^{2}<2.7\times 10^{-4}\mbox{ ,}\quad
\sum_{\alpha}|g_{\tau\alpha}|^{2}<5.5\times 10^{-2}.
\label{leptonb}
\end{eqnarray}

\subsection{Spectrum of lepton decay with Majorons}
\label{iic}
\begin{figure}[hbt]
\includegraphics[scale=0.27]{ElecSpec}
\includegraphics[scale=0.40]{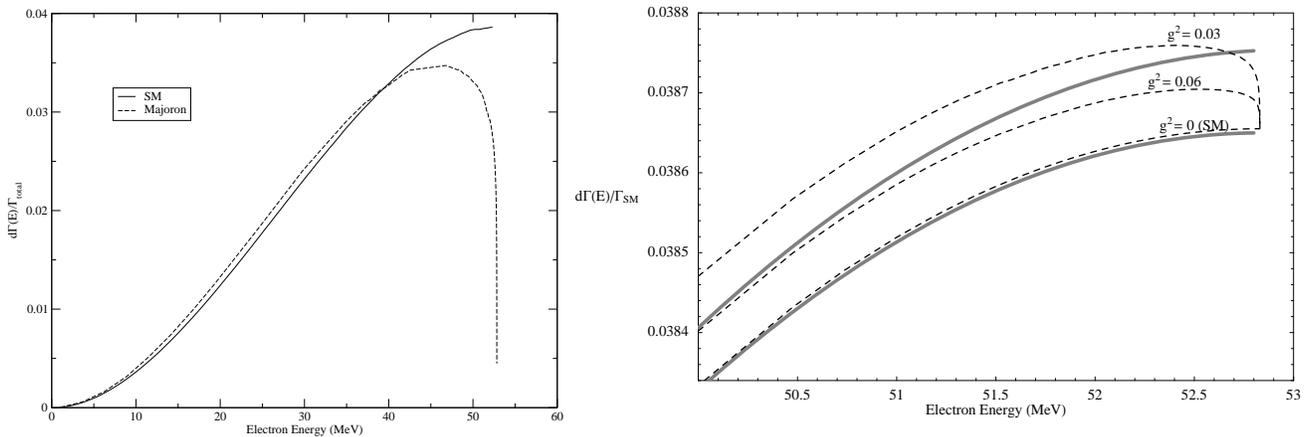}
\caption{At left, normalized electron spectra for muon decay in the SM
(solid line) and with Majoron emission only (dashed line). At right,
the experimental allowed region (between solid lines) for the electron
spectrum and the total predicted spectrum (SM plus Majorons) for three
values of $g^{2}=\sum_{\alpha}|g_{\mu\alpha}|^{2}$.}
\label{ElecSpec}
\end{figure}

Another method that can be used to improve the limits obtained above
is the analysis of the electron spectrum in the muon decay, which can be
modified by the inclusion of Majorons. The normalized spectrum for
the SM case and the Majoron case only are shown in
Figure~\ref{ElecSpec}.

Precision measurements of the electron spectrum (used to impose constraints on
non V-A interactions) may constrain $g$ if we consider the changes
in the SM spectrum after including Majoron emission. The usual analysis parametrizes
the electron spectrum using two parameters ($\rho$ and
$\eta$)~\cite{PDG}:
\begin{eqnarray}
\dfrac{d \Gamma
(x)}{dx}=\frac{G_{F}^{2}m_{\mu}^{5}}{48\pi^{3}}x^{2}[3(1-x)+
\dfrac{2}{3}\rho(4x-3)+3\eta\dfrac{m_{e}}{E_{max}}\dfrac{1-x}{x}]
\end{eqnarray}
where $x=\dfrac{E}{E_{max}}$ and
$E_{max}=\dfrac{m_{\mu}^{2}+m_{e}^{2}}{2m{}_{\mu}}$. For the SM the
predicted values are $\rho=0.75$ and $\eta=0$:
\begin{eqnarray}
\dfrac{d \Gamma_{SM}(x)}{dx}=\frac{G_{F}^{2}m_{\mu}^{5}}{48\pi^{3}}[\dfrac{3}{2}x^{2}-x^{3}]
\end{eqnarray}
The current experimental values are
$\rho=0.7509\pm0.001\mbox{ and }\eta=0.001\pm0.024$~\cite{PDG}.

When the total spectrum (SM plus Majoron) is considered, we have found
\begin{eqnarray}
\dfrac{d
\Gamma_{total}(x)}{dx}=\frac{G_{F}^{2} m_{\mu}^{5}}{48\pi^{3}}
[0.0066|g|^{2}-0.09|g|^{2}x
+(\dfrac{3}{2}+0.35|g|^{2})x^{2}-(1+0.25|g|^{2})x^{3}]
\end{eqnarray}
where $|g|^{2}=\sum_{\alpha}|g_{\mu\alpha}|^{2}$.

\begin{figure}[hbt]
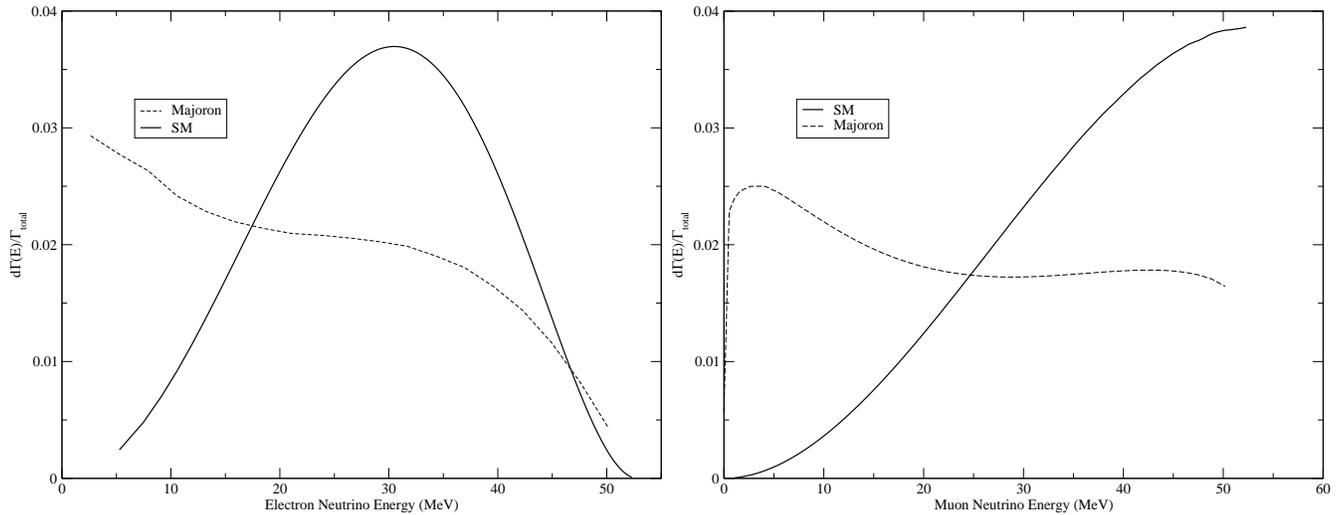

\includegraphics[scale=0.32]{EnSpec}
\includegraphics[scale=0.32]{MunSpec}
\caption{At left (at right), normalized electron neutrino (muon
neutrino ) spectra for muon decay in the SM is the solid curve and
with Majoron emission only is the dashed curve. In both cases we
assume a diagonal $g_{\alpha\beta}$.}
\label{EnSpec}
\end{figure}

From the above expression and Figure \ref{ElecSpec} we see that the
most sensitive region is at the end of the spectrum (large $x$),
which can be used to constrain $g$. Figure \ref{ElecSpec} also shows the
allowed region by experimental data (region between solid gray
lines) and the shape of the total spectrum (including the SM and
Majoron contributions) with different values of
$\sum_{\alpha}|g_{\mu\alpha}|^{2}$.
\begin{table}
\begin{tabular}{|c|c|}
\hline
Previous Bounds & Revised Bounds
\tabularnewline
\hline
\hline
&    \tabularnewline
$\sum_{\alpha}|g_{e\alpha}|^{2}<3\times10^{-5}$&
$\sum_{\alpha}|g_{e\alpha}|^{2}<5.5\times10^{-6}$
\tabularnewline
&    \tabularnewline
&    \tabularnewline  \hline
&    \tabularnewline
$\sum_{\alpha}|g_{\mu\alpha}|^{2}<2.4\times10^{-4}$&
$\sum_{\alpha}|g_{\mu\alpha}|^{2}<4.5\times10^{-5}$
\tabularnewline
&    \tabularnewline
&    \tabularnewline  \hline
&    \tabularnewline
none &
$\sum_{\alpha}|g_{\tau\alpha}|^{2}<5.5\times10^{-2}$
\tabularnewline
&    \tabularnewline
&    \tabularnewline  \hline
\end{tabular}
\caption{Comparison between the strongest bounds (including the scalar
$\chi$) obtained here and the previous bounds from the same
processes. All bounds are at 90\% C.L. and the previous bounds are
from ~\cite{Barger,Gelmini2,Britton:1993cj}.}
\label{FinalRes}
\end{table}

Because the spectrum is more sensitive to changes in the cubic term
(or the $\rho$ parameter), we consider the Majoron contributions to
$\rho$:
\begin{eqnarray}
\rho_{total}=\frac{3}{8}(2-0.25|g|^{2})
\end{eqnarray}
Using the chi-square method at 90\% C.L. we obtain:
\begin{eqnarray}
\sum_{\alpha}|g_{\mu\alpha}|^{2}<8\times10^{-3}
\label{leptonbound}
\end{eqnarray}

As can be seen in Figure \ref{EnSpec},
the Majoron main modifications to the spectra occurs
in the neutrino spectrum, which has been measured by the Karmen
experiment~\cite{Karmen}. However, due to experimental uncertainties,
the resulting bounds on $g$ are too weak in this case.

Summarizing, the strongest bounds are given in the Table
\ref{FinalRes}, where we compared the previous limits and the newest
constraints obtained here.

All bounds from Eqs.~(\ref{boundmeson}), (\ref{leptonb}),
(\ref{leptonbound}) can be written as
\begin{eqnarray}
\sum_{\alpha=e,\mu,\tau} |g_{l\alpha}|^{2} < L_{l}^{2}
\label{allbounds}
\end{eqnarray}
where $L_{l}^{2}$ is the strongest upper bound for $\sum_{\alpha}
|g_{l\alpha}|^{2}$ (see Table \ref{FinalRes}). From this
constraints, we assume the conservative limit, where the upper bound
applies not only for the sum, $\sum_{\alpha} |g_{l\alpha}|^{2}$, but
also for the individual elements, as $|g_{l\alpha}|$:
\begin{eqnarray}
|g_{l\alpha}|< L_{l},\;\forall\;\alpha=e,\mu,\tau
\end{eqnarray}

\subsection{Mass Basis}
\label{iid}
All the results obtained so far are written in the flavor basis. However,
in many cases, theoretical analysis are easier on the mass basis.
We have two possible cases: Dirac or Majorana neutrinos. In this
section we assume Majorana neutrinos to transform our bounds to the
the mass basis.

We can translate the previous results to the mass basis using
the transformation matrix $U$~\cite{PDG}:
\begin{eqnarray}
U=\left(\begin{array}{ccc}
c_{12}c_{13} & s_{12}c_{13} & s_{13}e^{-i\delta}\\
-s_{12}c_{23}-c_{12}s_{23}s_{13}e^{i\delta} &
c_{12}c_{23}-s_{12}s_{23}s_{13}e^{i\delta} & s_{23}c_{13}\\
s_{12}s_{23}-c_{12}c_{23}s_{13}e^{i\delta} &
-c_{12}s_{23}-s_{12}c_{23}s_{13}e^{i\delta} &
c_{23}c_{13}\end{array}\right)\times
diag(e^{i\alpha_{1}/2},e^{i\alpha_{2}/2},1)
\end{eqnarray}
where $c_{ij}=cos(\theta_{ij})$ and $s_{ij}=sin(\theta_{ij})$. The
neutrino mass matrix is given by $M=diag(m_{1},m_{2},m_{3})$ and for a
given mass $m_1$, we can written all other masses as a function of
$m_1$ and the squared mass differences as follows
\begin{eqnarray}
\Delta m_{12}^{2}\equiv m_{2}^{2}-m_{1}^{2}=\Delta m_{\myodot}^{2}
\mbox{ and }
\Delta m_{23}^{2}\equiv m_{3}^{2}-m_{2}^{2}=\Delta m_{atm}^{2}.
\end{eqnarray}
Although the mass differences and angles have
been measured experimentally~\cite{PDG},
we have no information on the Majorana phases $\delta$, $\alpha_{1}$ and
$\alpha_{2}$.
\begin{figure}[hbt]
\includegraphics[scale=0.47]{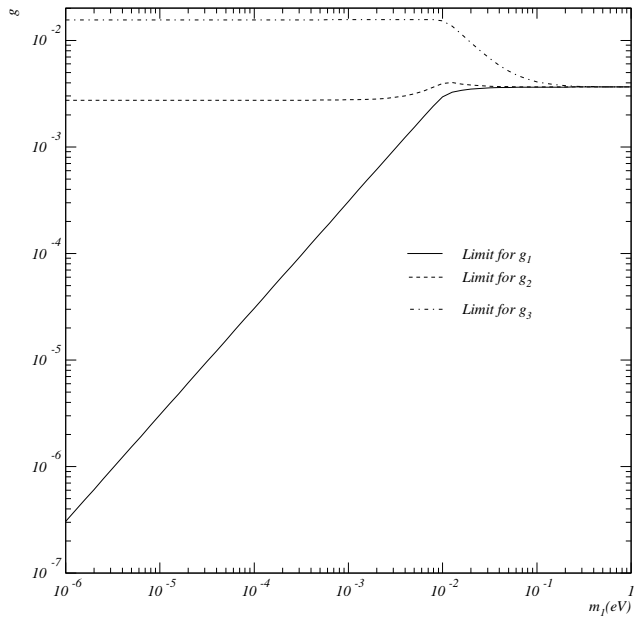}
\caption{Exclusion curves in the $m_{1}-G$ plane. We show the curves
for $g_1$, $g_2$ and $g_3$. These curves are for $\theta _{13}=0$,
non-zero values of $\theta_{13}$ are more restrictive.}
\label{MassBounds}
\end{figure}
To calculate the bounds in the mass basis, we will use the transformation rule for
Majorana neutrinos
\begin{equation}
G=U^{T}gU
\label{eq2}
\end{equation}
where g is the neutrino-Majoron coupling matrix in the flavor basis
and G is the neutrino-Majoron coupling matrix in the mass basis.

Although it is not valid in general, many models~\cite{Chikashige:1980ui,Joshipura:1992hp,Schechter:1981bd} have
the following  property (at least in some limit)
\begin{equation}
G=diag(g_{1},g_{2},g_{3})\propto M=
diag(m_{1},m_{2},m_{3})
\label{eq1}
\end{equation}

Following~\cite{tomas1}, we calculate the allowed region for
different values of $\delta$, $\alpha_{1}$ and $\alpha_{2}$ and then
choose the union of these regions as the final result, valid for any
value of the phases, as shown  in Figure \ref{MassBounds}.

\section{Conclusions}
\label{iii}
Using three different techniques we were able to
constraint the neutrino-Majoron couplings. The strongest constraints
are shown in Table \ref{FinalRes}. Considering only the limits
from meson decays we improve by one order of magnitude the previous
limits on $|g_{e\alpha}|^{2}$ and
$|g_{\mu\alpha}|^{2}$~\cite{Barger,Gelmini2,Britton:1993cj}.
Although the best constraints were obtained from meson decay rates,
we have shown that independent bounds can also be obtained from
$\mu$ and $\tau$ decays. The latter one being the best to constraint
the $g_{\tau\alpha}$ elements. We stress that the bounds on
$g_{\tau\alpha}$ shown in Table \ref{FinalRes} is probably the first
model-independent constraint for this parameter.

The third alternative used was an analysis of the spectrum of muon
decay. Despite its potential for constraining the $g_{\mu\alpha}$
elements, the experimental values are not precise enough to make such
an analysis useful.  Our best constraints  are
$
|g_{e\alpha}|^{2}<5.5\times10^{-6}$,
$|g_{\mu\alpha}|^{2}<4.5\times10^{-5}$ and
$
|g_{\tau\alpha}|^{2}<5.5\times10^{-2}$, $\alpha=e,\mu,\tau$ at 90~\%~C. L. .

Because the models cited here usually try to explain the neutrino
mass scale, it may be convenient to analyze the limits on
neutrino-Majoron couplings in the mass basis. With that in mind we
transformed all our results from the flavor basis to the mass basis,
using the current values for the angles of the neutrino mixing
matrix. As shown in Figure \ref{MassBounds} the constraints on the
mass basis are usually weaker than those on the flavor basis.

\begin{acknowledgments}
This work was supported by Funda\c{c}\~ao de Amparo
 \`a Pesquisa do Estado de S\~ao Paulo (FAPESP) and  Conselho
Nacional de Desenvolvimento Cient\'\i fico e Tecnol\'ogico
(CNPq).
\end{acknowledgments}


\end{document}